\documentclass[preprint,prl,showpacs,aps]{revtex4}
\begin{document}

\title
{Euler buckling induced folding and rotation\\of red blood cells in an 
optical trap}
\author{A. Ghosh$^{a,b}$, Supurna Sinha$^{a,b}$, J. A. Dharmadhikari$^c$, 
S. Roy$^c$, A.~K.~Dharmadhikari$^c$, J. Samuel$^{a,b}$, S. Sharma$^c$, and D. Mathur$^c$}
\address{$^a$Harish Chandra Research Institute, Chhatnag Road, Allahabad 211 019, India}
\address{$^b$Raman Research Institute, Bangalore 560 080, India}
\address{$^c$Tata Institute of Fundamental Research, 1 Homi Bhabha Road, Mumbai 400 005, India.}
\date{\today}

\begin{abstract}

We investigate the physics of an optically-driven micromotor of biological origin. A single, live red blood cell, 
when placed in an optical trap folds into a rod-like shape. If the trapping laser beam is circularly 
polarized, the folded RBC rotates. A model based on the concept of buckling instabilities 
captures the folding phenomenon; the rotation of the cell is simply understood using the 
Poincar\`e sphere. Our model predicts that (i) at a critical intensity of the trapping beam the RBC shape undergoes
large fluctuations and (ii) the torque is proportional to the intensity of the laser beam. These 
predictions have been tested experimentally. We suggest a possible mechanism for emergence of birefringent properties in the RBC in the folded state.   
\end{abstract}

\maketitle

\section{Introduction}

Micromanipulation of biological matter using light is of great current interest because of its relevance to fundamental research as well as applications. Confinement of a single cell in an optical trap is one recent example of how optically-generated forces affect cellular dynamics. Recent advances have enhanced the capability of applying and sensing forces and displacements with magnitudes in the picoNewtons and nanometer ranges, respectively, and with corresponding sensitivities of femtoNewtons and sub-nanometers. Optical forces that lie in the range 1-50 pN are capable of physically deforming a cell without causing cell death: this has been demonstrated \cite{Svoboda} in trapping experiments involving single red blood cells (RBCs).

We have recently conducted experiments to investigate the behavior of live human red blood cells under optical forces generated using 
both linearly and circularly polarized light \cite{cs, apl}. These experiments have shown that a normal human RBC, 
which has a biconcave disk shape, approximately 8 $\mu$m in diameter, deforms into a folded shape upon being placed in an optical trap; the trapped RBC subsequently rotates when circularly polarized laser radiation is used. The rotational speed is controlled by the magnitude of the laser power that is applied. Experiments carried out using RBCs from mice, 
which have a range of diameters (4-8 $\mu$m), show that the rotation speed also depends upon cell size \cite{apl}.
   
This paper is devoted to understanding the physics behind the folding and  rotation of RBCs. We present here the results of a study on the optically-induced folding and rotation of a red blood cell kept under buffered conditions that are close to physiological conditions. Experiments that involve optically-driven processes and devices carry a distinct advantage since forces can be applied without mechanical contact. Consequently, there has been a resurgence of interest in optically-driven micro-motors. Following the classic seven-decade-old experiment of Beth \cite{beth}, which demonstrated the conversion of optical energy into mechanical energy, leading to rotation of micron-sized quartz crystals in circularly polarized light, other non-contact modes of rotation have been demonstrated in micro-structures comprising specially shaped dielectrics and birefringent particles \cite{higurashi,galajda, friese}. While the rotation induced by optical forces can be finely controlled, because it depends on parameters like incident laser power and polarization state, in hitherto existing work the stringent constraint on the shape and microfabrication of rotors continues to be a key challenge. 
Thus, experiments involving optical trapping of {\it naturally occurring} material, like a live red blood cell (RBC) \cite{cs,apl} or other single living cells \cite{chlamy} is of great relevance since these obviate the need for microfabrication of special shapes. 

The collapse of structures under compressive mechanical stress, or under their own weight, is of wide interest 
in many fields \cite{buckling} and such instabilities have been studied in the classical elasticity 
of rods and plates \cite{buckling, landau_lifshitz}. When a structure is subjected to compression it undergoes large displacements transverse to the load, and buckles, as can be readily demonstrated with a drinking straw or a ball pen refill. The buckling process is relevant not only to macroscopic systems described by classical elasticity, but also to 
microscopic objects like semiflexible polymers, biological membranes and metallic films.  We notice here that the Euler instability also affects a red blood cell placed under the delicate compressive forces of an optical trap \cite{freya}.  

The paper is organized as follows. We first summarize our experimental observations. After presenting a brief overview, from a physics perspective, of the structural properties of RBCs, we present a simple model that captures the observations that we have made in our experiments. The subsequent section deals with the predictions made by our theoretical model. 
We end with some concluding remarks. 

\section{Observations}

Our measurements on live RBCs were conducted using a single-beam optical trap \cite{cs} comprising a 1064 nm, 1 W diode pumped Nd:YVO$_4$ laser whose light was sharply focused with a 100X oil-immersed objective of large numerical 
aperture (NA=1.3). A red blood cell under physiological pH conditions was optically trapped within the focal volume of the beam. The cell was observed to fold when subjected to a trapping force of the order of 10 pN (corresponding to laser power of $\sim$20 mW). Upon removal of the optical trap (blocking the laser beam), the RBC was observed to unfold to its original biconcave shape. When circularly polarized laser light was used, the folded RBC was observed to rotate \cite{apl}. The experimental observations of folding can be summarized as follows (see Fig. $1$):

\begin{enumerate}
\item The RBC folds in the optical trap and unfolds when the trap is off. It is observed that the cell {\it folds} and does not {\it flip} as can be checked from the fact that the folded RBC exhibits a width ($\sim$3.8 $\mu$m) that is almost twice that of the thickest part of the unfolded cell. 
\item The folding time (typically $\sim1 s$) is much smaller than the unfolding time (typically $\sim14 s$).
\item If the incident beam is linearly polarized, the long axis of the folded RBC aligns itself in the direction of the electric field of the incident trapping laser beam and faithfully follows the electric vector as the polarization is rotated.
\end{enumerate}

The experimental observations pertaining to RBC rotation can be summarized as follows:

\begin{enumerate}
\item The folded RBC rotates in circularly polarized light but not in linearly polarized light. The sense of rotation depends on the sense of circular polarization.
\item The rotational speed has been shown to be controlled by the magnitude of the laser power that is applied. 
\end{enumerate}

\section{Structure of RBCs: an overview} 

The standard picture of a red blood cell is that of a semi inflated bag containing a viscoelastic incompressible fluid, the cytoplasm. The bag is a plasma membrane consisting of a phospholipid bilayer containing other macromolecules, which form the cytoskeleton. The cytoskeleton is a protein network whose links are spectrin filaments (length $\sim$200 nm) meeting at junctions of short actin filaments (length $\sim$37 nm). Other proteins like ankyrin bind the mesh as a whole to
the cytoplasmic side of the phospholipid bilayer. This protein network is important in determining the elastic and mechanical properties of the cell which are characterized by the elastic constants of the membrane. In particular, the elastic modulus for area compressibility $K$ is in the range $\sim$3-6 $\times$ 10$^{5}$ pN $\mu$m$^{-1}$, the elastic modulus for shear $S$ is  $\sim$5-7 pN $\mu$m$^{-1}$ and the bending elastic modulus $B$ is $\sim$10$^{-1}$ pN $\mu$m \cite{jerusalem}. For displacements in the micron range, it is clear that the elastic energies of bend, shear and compression are in the ratio $1:50:10^{6}$, that is, 
$$ E_{bend} \ll E_{shear} \ll E_{comp}.$$
In other words, it is easier to bend the cell than to shear it and it is hardest to compress it. A detailed analysis of the cell shape and elasticity of the red blood cell is presented in \cite{cyto}. Our focus here is to study a particular aspect of RBC elasticity which involves buckling of the cell in the presence of light forces. 

The elastic properties of the RBC are biologically crucial as they permit it to squeeze through narrow capillaries.
The natural unstressed shape of the RBC is determined by the elastic properties of the membrane, its area and the enclosing volume. Under the action of external forces, the red cell deforms but recovers its original shape when the forces are removed. The bilayer provides resistance to bending and the cytoskeleton resists shear as well as compression. Earlier studies of RBC elasticity have been carried out using techniques like micro-pipette aspiration, laminar shear flow, and optical stretchers. We reiterate that the present study focuses on live cells, maintained under physiological conditions. Our experiments are distinct from a recent work on the exertion of optical forces on cells that are no longer live by virtue of the conditions of the medium in which they are maintained \cite{cat}.

\section{Our simplified model}

We find that the main experimental observations can be adequately captured by a simple model which is independent of the detailed structure of the RBC. What is appealing about the model is its simplicity, that it reproduces the main experimental findings and the fact that it makes testable predictions.

\subsection{Cell folding}

In our simplified model we neglect the biconcave geometry of the RBC and consider it to be a flat disk shaped elastic membrane with an energy cost for bending and deformation. Our model makes some testable predictions.
 
The elastic energy for the RBC disk has two terms \cite{kantor}: (i) the energy cost associated with the extrinsic (mean) 
curvature of the membrane, $E_c=B \int d^2 \sigma (H)^2$, where $d^2 \sigma$ represents an area element, $B$ is the bending modulus, $H$ is the membrane's extrinsic curvature, and (ii) the energy cost of the strain internal to the membrane (shear and compression) that is given by an integral over the membrane of the square of the strain tensor and is characterized by the Lam\'e constants \cite{landau_lifshitz}. So, an unstrained RBC is a disk with zero intrinsic and extrinsic curvatures. We confine ourselves to the lowest energy description of the system in which we neglect the effects of shear and compression and describe the system purely in terms of the bending energy. The membrane only assumes shapes with zero intrinsic curvature because it takes far greater energy to change the intrinsic geometry. Apart from the membrane's elastic energy there is also energy associated with the optical trap, 
\begin{equation}
E_{\rm trap}=-\frac{t}{2}\int d^2 \sigma \epsilon I(x,y), 
\end{equation}
where $t$ is the thickness of the membrane, 
$\epsilon$ is the dielectric constant of the RBC and 
$I(x,y)$ is the local light intensity in the focal plane. 
The higher the intensity, the lower is the potential 
\begin{equation}
V(x,y)=-\frac{1}{2} t \epsilon I(x,y).
\end{equation}
We expand the potential in a Taylor series around the minimum of the potential ($z$ is chosen along the direction of propagation of light),  
\begin{equation}
V(x,y)= {1\over{2}} A (x^2+y^2),
\end{equation}
where $A$ is a constant proportional to the incident laser power. $E_c$ is lowest when the RBC is flat and unfolded. $E_{\rm trap}$ is lowest when the RBC is at the bottom of the potential. The shape of the membrane is determined by a competition between $E_c$ and $E_{\rm trap}$. 

So, when the trap is switched on, for high enough laser power the trap energy overcomes the elastic energies of the membrane 
and folds it, and when the laser trap is switched off, the flat state is more favorable. The constraint of preserving the intrinsic geometry forces the membrane to assume a cylinder-like shape \cite{foot} (Fig. 1) rather than any other shape that has non-zero intrinsic curvature. The allowed configurations of the elastic membrane are similar to the allowed configurations of a disk made of paper: 
it can be bent into a cylinder, but not into a sphere as this deforms its internal geometry. An analysis of the two competing energy terms shows that the threshold power needed for folding is inversely proportional to the cube of the linear dimension of the cell. This is a prediction of the model which can be tested against future experiments on this system. 

As noted, we view the folding of the trapped RBC as an analog of the buckling instability \cite{landau_lifshitz,freya}. Although the RBC is modeled as a $2D$ disk, it bends only in one direction because of the constraint of preserving intrinsic geometry and so the problem can be treated as essentially one dimensional, similar to a compressed rod. We 
simplify the analysis by focusing on the lowest energy mode of deformation, in which the RBC assumes a shape which is part of a cylinder of radius $R$. We use as the order parameter $\alpha=d/R$, where $d$ is the RBC diameter. $\alpha=0$ corresponds to $R=\infty$, the unfolded state; a non-zero $\alpha$ describes a curved configuration. We would 
expect the energy of bending to be proportional to $\alpha^2$, since bending in either direction costs the same amount of 
energy, $E_c=a \alpha^2/2$, where $a$ is a constant related to the bending modulus. 
Similarly, the trap energy is also symmetric in $\alpha$ and can be expressed as $E_{trap}=b \alpha^2/2$, where $b$ is a constant related to the intensity of the laser radiation. The competition between $E_c$ and $E_{trap}$ determines the stability of the folded state.

\subsection{Cell rotation}

We now consider the rotation of a trapped RBC. From the observation that the long axis of the RBC always follows the 
polarization vector one can infer that the folded RBC is birefringent. Thus it follows that in the presence of circularly polarized light the folded RBC will rotate. While this can be understood using the Maxwell stress tensor of the 
classical electromagnetic field \cite{born}, we note here that the same results can be obtained in a more transparent way by invoking the spin of the photon. In passing through the folded, birefringent RBC, the light polarization changes from circular to elliptical. The difference in spin angular momentum is imparted to the RBC, thus exerting a torque on it causing it to rotate. The photon ``spin" is best described by the Poincar\`e sphere where for spin=1$\hbar$, the north pole ($\theta=0^\circ$) represents right circularly polarized light with angular momentum $\hbar$ while the south pole $(\theta =180^\circ $) represents left circularly polarized light, with angular momentum $-\hbar$. Linearly polarized light carries no
spin angular momentum and is represented along the equator while all other points on the sphere represent elliptically polarized light.

The change in angular momentum caused by a single photon is equal to the change in $\cos\theta$ and, therefore, to the change in the z-coordinate of the Poincar\'e sphere, multiplied by the unit $\hbar$ of angular momentum. One can express the torque as the total angular momentum transferred per unit time: $\tau=\gamma N \hbar=\gamma P/\omega$, 
where $\gamma = 1 - cos[2\pi t (n_1-n_2)/\lambda]$ is a dimensionless number, the change in the $z$ coordinate on the Poincar\'e sphere (here $t$ is the thickness of the sample, $n_1$ and $n_2$ are the refractive indices of the ordinary and the extraordinary ray respectively and $\lambda$ is the wavelength of light used), N is the number of photons per second, and $P$ is the power of the incident light (number of photons per second multiplied by the energy $\hbar \omega$ of each photon). We, therefore, expect the torque that is generated to be proportional to the incident laser power as is,
indeed, observed in our measurements (Fig. 2). 

Circularly polarized light exerts a torque on the cell which causes it to rotate in the surrounding viscous medium. Since we 
are in the low ($<<$ 1.0) Reynold's number regime we would expect the angular velocity $\omega$ to be proportional to the torque $\tau$:
\begin{equation}
\tau=  \xi\omega, 
\end{equation} 
where $\xi$ is the frictional drag coefficient given by
\begin{equation}
\xi = \frac{(\pi/3) \eta L^3}{\ln(\frac{L}{2r}) - 0.447}.
\end{equation} 
 
From the observed angular velocity we calculate the torque to be large, 26 nN nm, using the value L = 7 $\mu$m for the length of the cylinder, r=2 $\mu$m for the radius of the cylinder and $\eta$ = 0.0013 Nsm$^{-2}$ for the viscosity of the surrounding liquid.

\section{Predictions}

The time scales for folding and unfolding can also be understood in our model with two parameters: the depth of the trap that is represented by $b$ (experimentally controlled by the light intensity), and the membrane stiffness represented by $a$. When $\alpha$ deviates from $0$, it experiences a restoring force, $F=(a-b)\alpha$. Note that all measurements were made at low Reynolds number where viscous forces dominate over the negligible \cite{berg} inertial forces. As in any viscous medium, the force $F$ causes a motion whose ``velocity"  $\dot{\alpha}$ is proportional to the applied force $\dot{\alpha}\propto (a-b) \alpha$. From this we conclude that the timescale for folding is given by $T_{\rm fold}=1/(a-b)$, 
or in terms of the folding rate, $1/T_{\rm fold}\propto P-P_0$, where $P_0$ is the critical laser power for folding. 
Since velocities are directly proportional to forces, timescales are inversely proportional to forces. 
The higher the force, the smaller the timescale. This picture explains why the folding time is much shorter than the unfolding time. The trap force induces folding and the process is fast. On turning the trap off, the membrane spontaneously 
and slowly relaxes back to its flat state as there are no forces to keep it folded. Our model predicts that 
by controlling the light intensity one can change the folding time. As the laser power is lowered, the folding time is expected to increase and at a certain low power, when the trap energy just cancels the elastic energy, we expect to see large fluctuations in the shape of the membrane. This is because at this power level there is no 
restoring force in the $\alpha$ variable. The membrane is floppy and susceptible to Brownian fluctuations\cite{freyb}. 
These large shape fluctuations have been experimentally seen. 

Measurements of the dependence of the folding rate on normalized laser power are shown in Fig.3; our results are in accord with the predictions of our model.

\section{Conclusions}
 
We have presented the results of a combined experimental and theoretical study of folding and rotation of a RBC in an optical trap. The elastic properties of the RBC are a sensitive function of the pH of the buffer solution. We emphasize that our experiments were carried out under physiological conditions. The main observation is that a single red blood cell, when placed in an optical trap folds into a rod-like shape; if the trapping laser beam is circularly polarized, the folded RBC rotates. We have proposed a simple model based on the notion of Euler buckling instability to capture the physics of folding. The rotation of the cell is understood in terms of a simple picture involving the Poincar\`e sphere. The predictions that emerge from the model have been successfully tested experimentally.

We notice that the orders of magnitude of the torques achieved in these experiments are significantly higher than those achieved in quartz crystals. This suggests the following model for the appearance of birefringence in a 
cylindrically folded RBC. In the folded state, the long biomolecules constituting the cytoskeleton of the RBC, 
which are aligned along the axis of the cylinder contribute to birefringence somewhat like the very long organic molecules lined up in a liquid crystal to give it birefringent properties. In the unfolded state the cytoskeletal biopolymers are randomly oriented in the plane transverse to the incident beam and thus do not cause birefringence. Further investigations are needed to confirm this model of birefringence.

\newpage

\begin{figure}
\caption{(Color online) On placing a biconcave, disk-like red blood cell in an optical trap, it undergoes folding that is depicted in (a) frames from a real-time movie and (b) in cartoon format. Each movie frame is taken 250 ms after the preceding one. Upon removal of the trap, unfolding occurs over a typical timescale of {\it ca.} 14s.}
\end{figure}

\begin{figure}
\caption{Variation of the torque generated in a rotating RBC with incident laser power, using circularly polarized light.}
\end{figure}

\begin{figure}
\caption{Dependence of folding rate on normalized laser power, $P-P_0$, where $P$ is the incident laser power and $P_0$ is the power (4 mW) below which folding does not occur for a 6$\mu$m RBC.}
\end{figure}


\begin{references}
\bibitem{foot} Notice that we restrict ourselves to the lowest energy description that would dominate the physics of buckling. One can imagine other structures, for instance, a corrugated sheet that would be consistent with the constraint of fixed intrinsic geometry but such a structure would pertain to a higher energy and is neglected in the present analysis. The experimental observations also support our low energy description.
\end{references}
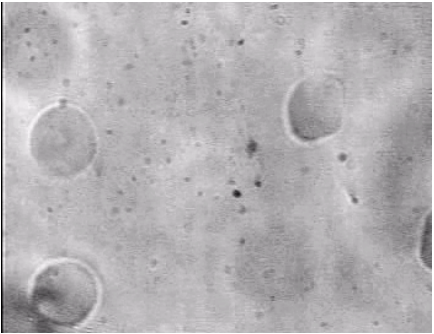
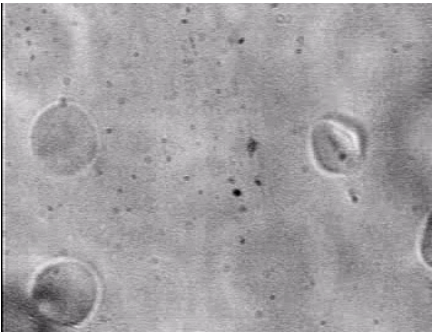
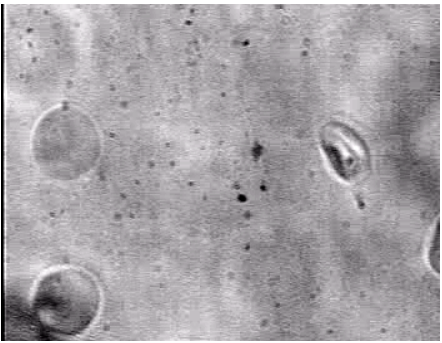
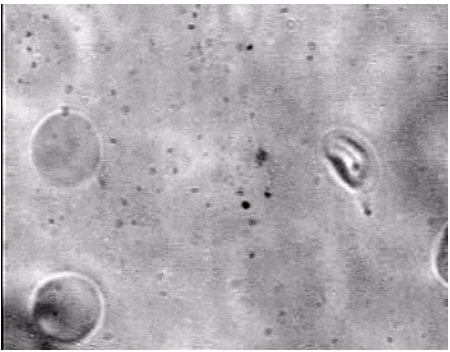
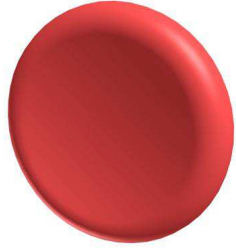
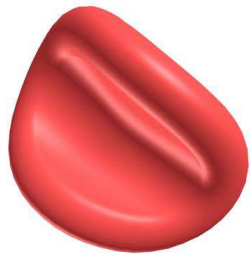
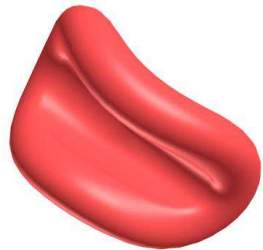
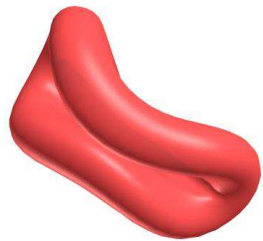
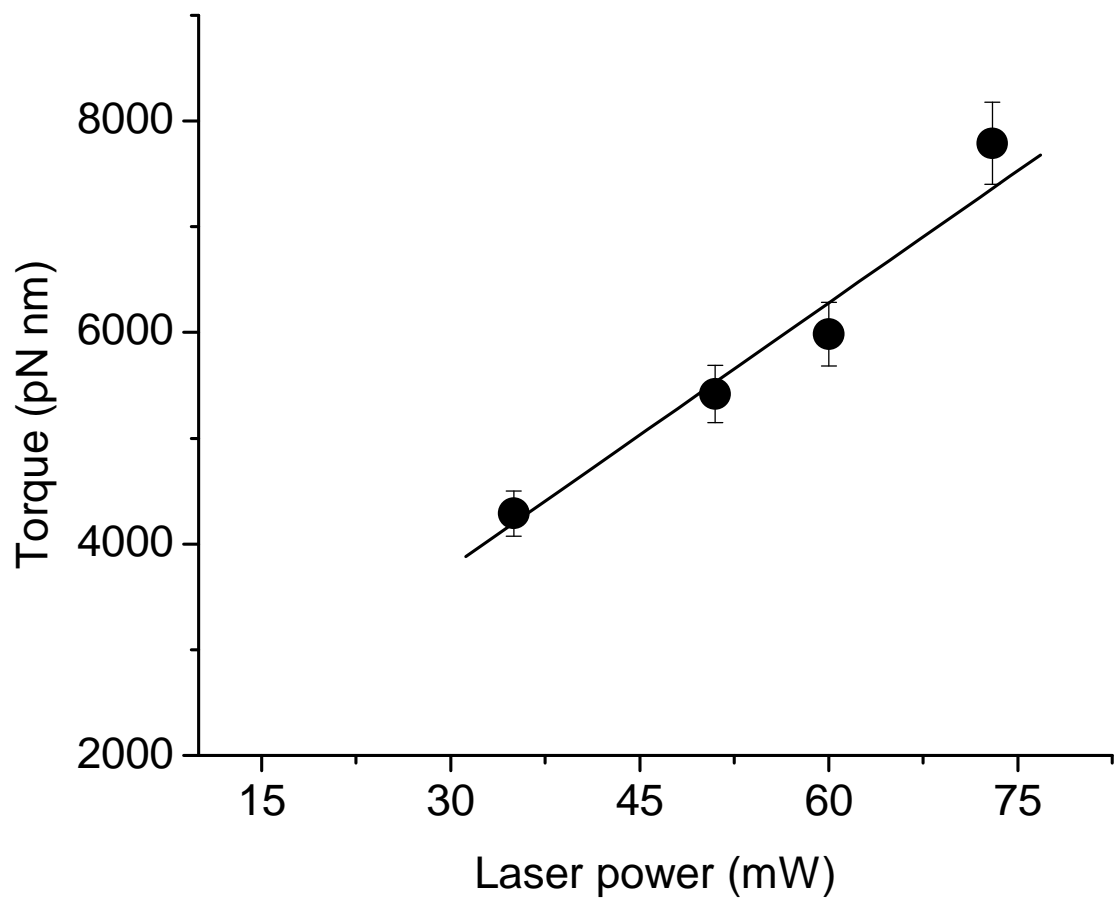
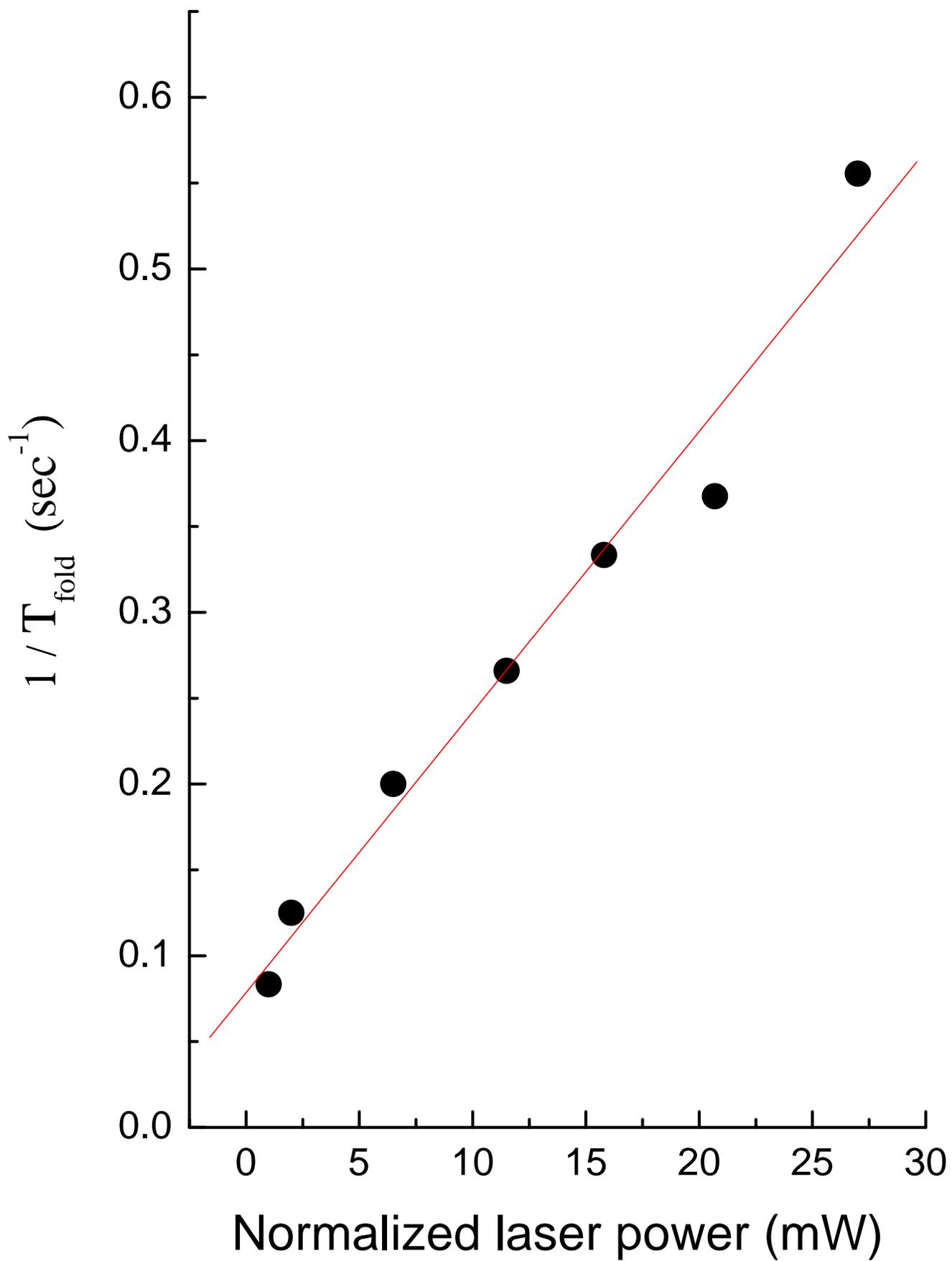
\end{document}